\documentclass[%
 reprint,
 amsmath,amssymb,
 aps,
]{revtex4-2}

\usepackage{graphicx}
\usepackage{dcolumn}
\usepackage{bm}
\usepackage[utf8]{inputenc}
\usepackage[T1]{fontenc}
\usepackage{mathptmx}
\usepackage{subfigure}

\begin{document}

\preprint{APS/123-QED}

\title{Actuation system for inertial sensors in high-precision space missions using a torsion pendulum }

\author{Fangchao Yang}
 \email{yangfc37@zhejianglab.com}
\author{Yan Zhu}
\author{Xiaofei Jin}
\affiliation{
Zhejiang Lab, Hangzhou 311121, People’s Republic of China.
}%
\author{Yujie Zhao, Wei Hong}
 \email{hongwei@hust.edu.cn}
\affiliation{%
MOE Key Laboratory of Fundamental Physical Quantities Measurement, Hubei Key Laboratory of Gravitation and Quantum Physics, PGMF and School of Physics, Huazhong University of Science and Technology, Wuhan 430074, People's Republic of China.
}%
\author{Shixun Pei}
\affiliation{%
School of Aero-Engine Engineering, Zhengzhou University of Aeronautics, Zhengzhou 450046, People’s Republic of China.
}%



\date{\today}

\begin{abstract}
Precision space inertial sensors are imperative for Earth geodesy missions, gravitational wave observations, and fundamental physics experiments in space. In these missions, the residual acceleration noise of a test mass (TM) caused by the forces from inertial sensor components and the environment must be maintained below a specific level. Several forces contributing to residual acceleration are related to the actuation system; hence, developing a precise actuation system is crucial to eliminate erroneous forces and achieve highly accurate measurements of TM acceleration noise. However, obtaining a “free-fall” status of the TM to make sensitive ground tests for actuation system poses significant challenges. In this study, a torsion pendulum was employed to analyze the impact of the actuation system on TM torque noise. A closed-loop control system integrating the torsion pendulum with components of the actuation modules was designed to assess the performance of the actuation control algorithm. Experimental findings indicate that certain parameters such as the maximum torques and DC bias voltages in an actuation system contribute to additional torque noise, with the peak noise level reaching up to $10^{-{13}}$ Nm/{Hz}$^{1/2}$ at 1 mHz. The closed-loop system demonstrates a stable tracking error of approximately 10$^{-7}$, indicating the robust tracking performance and robustness of the integrated system for TM rotation control in varying inertial sensor conditions. The proposed method serves as a reference for advancing actuation systems in diverse space gravitational missions.
\end{abstract}

\maketitle
\newcommand{\RNum}[1]{\uppercase\expandafter{\romannumeral #1\relax}}

\section{Introduction}
High-precision space-based gravitational experiments employ inertial sensors to accurately measure the relative motion between a free-fall test mass (TM) and the spacecraft; these experiments include tests of fundamental physics\cite{STEP, Everitt2015, Touboul2017}, studies of the Earth's geopotential\cite{GOCE2010, Grace2019}, and gravitational wave observatories\cite{LISA, TAIJI, TIANQIN, DECIGO2010}. Typically, these sensors comprise a TM in near free-fall, enclosed in an electrode housing (EH) containing an array of electrodes. These electrodes can measure the non-conservative force on the spacecraft or serve as part of a laser interferometry system for inertial reference. The electrostatic actuation system, a touch-free mechanism, is critical in capacitive sensing and controlling the position of the free-floating TM relative to the EH in space\cite{Armano_sensing, LPF}.

In various space-based experiments, the spacecraft slowly drifts against the free falling TM because of the impact of particles coming from the sun (the solar wind) or from outer space\cite{Mance2012}. The micro-Newton thrusters on spacecraft compensate for this drift, but not in all degrees of freedom (DOFs)\cite{Armano2019}. Therefore, electrostatic actuation must be used for those DOFs that are not controlled by the spacecraft thrusters to correct the position of the TM relative to the spacecraft. To achieve the target of these space-based gravitational missions, an inertial sensor with ultralow residual acceleration (less than ${10^{-16}}$ m/s$^{2}$/{Hz}$^{1/2}$) is required. Actuation noise from electrostatic actuation is one of main sources of the TM residual acceleration at frequencies below 1 mHz\cite{Castelli2020, LPF}. Therefore, investigation of the influence of the actuation system on the TM acceleration noise and verification of the actuation control algorithm is necessary and essential for most space missions.

Studies have been conducted on the electrostatic actuation system. In \cite{Armano2020_actuation}, the accuracy of the electrostatic actuation system and its impact on the LiSA Pathfinder (LPF) main observable were investigated. The results showed that inaccuracy is caused mainly by rounding errors in waveform processing and by the random error due to the analog-to-digital converter and random noise in the control loop. However, other factors affecting the TM residual acceleration were not studied. In \cite{Zhang2023}, an actuator circuit was designed and a low-cost hardware-in-the-loop simulation system combining the actuation circuit and other parts of the inertial sensor was established to analyze the residual acceleration noise. However, the inertial sensor was a simulation model rather than an experimental system. Therefore, developing a ground test facility to investigate the TM control performance and the influence of the actuation system on TM residual acceleration for sensitive space-based gravitational experiments remains a crucial and vital challenge.

In this study, a torsion pendulum was constructed to investigate the influence factors in an actuation system. A combined control system including the torsion pendulum and actuation modules was designed to verify the actuation control algorithm and evaluate tracking performance and robustness for TM rotation control. The structure of this paper is as follows. First, the theoretical derivation of the TM actuation algorithm and the influence factors on the TM torque noise are described. Then, our experimental setup is introduced. The results of tests with different maximum torques and DC bias voltages are then described, and the control performance of the TM rotation is discussed. Finally, the conclusion and discussion are presented.

The contributions of this study are as follows: (a) Development of a ground test facility based on a torsion pendulum to evaluate the influence of an actuation system on TM torque noise, and (b) design of a closed-loop control system to verify the actuation algorithm and evaluate the tracking performance and robustness for TM control.

\section{Modeling and Analysis}

\subsection{\label{sec:level1} TM Actuation Algorithm}
The inertial sensor comprised a TM, six pairs of sensing and actuation electrodes, and three additional pairs of electrodes responsible for injecting a high-frequency voltage on the TM to enable electrostatic sensing. To determine the required actuation voltages for specific forces and torques, several assumptions and simplifications were made based on the actuation scheme used by the LPF\cite{actuation_tech}: (1) All the electrodes apply actuation voltages except the injection electrodes; hence, the voltages on the injection electrodes are zero; (2) the TM charge is negligible because the charge management system controls the charge below a certain level; (3) all electrodes are treated as ideal planar capacitors; hence, any force or torque contribution coming from other directions are considered negligible; (4) the sum of the voltages on the electrodes must maintain zero TM potential; hence, the induced voltage on the TM generated by the actuation voltages is maintained at zero.

By simplifying the inertial sensor as a system of $N$ conducting surfaces, including $N – 1$ electrodes and other surfaces on the housing and the TM, the electrostatic force can be expressed as\cite{Yang2020}
\begin{equation}
{F_{\rm{k}}} = \frac{1}{2}\sum\limits_{i = 1}^{N - 1} {\frac{{\partial {C_{{\rm{i,TM}}}}}}{{\partial k}}} {({V_{\rm{i}}} - {V_{\rm{T}}})^2} + \frac{{{Q_{{\rm{TM}}}}^2}}{{2{C_{\rm{T}}}^2}}\frac{{\partial {C_{\rm{T}}}}}{{\partial k}} - \frac{{\partial {V_{\rm{T}}}}}{{\partial k}}{Q_{{\rm{TM}}}}
\label{eq:1},
\end{equation}
where $k$ is a generalized coordinate for a DOF, $V_{\rm{i}}$ is the potentials of the electrode surfaces, ${C_{{\rm{i,TM}}}}$ is the capacitance between the $i$th surface and TM, ${Q_{{\rm{TM}}}}$ is the free charge accumulated on the TM, and ${V_{\rm{T}}}$ is the potential induced by voltages on the other conductors in the entire sensor. Under these assumptions and simplifications, equation~(\ref{eq:1}) can be simplified as
\begin{equation}
{F_{\rm{k}}} = \frac{1}{2}\sum\limits_{i = 1}^{N - 1} {\frac{{\partial {C_{{\rm{i,TM}}}}}}{{\partial k}}} {V_{\rm{i}}}^2
\label{eq:2}.
\end{equation}

In this study, a torsion pendulum was utilized to simulate the rotational motion of a TM. Hence, we are concerned only with the electrostatic torque around the $\varphi$ direction; thus, equation ~(\ref{eq:2}) can be rewritten as
\begin{equation}
{N_{\rm{\varphi }}} = \frac{1}{2}\sum\limits_{i = 1}^4 {\frac{{\partial {C_{{\rm{i,TM}}}}}}{{\partial \varphi }}} {V_{\rm{i}}}^2
\label{eq:3}.
\end{equation}
When the TM is controlled in high-resolution mode, the actuation stiffness is assumed to be constant. It can be shown that the approximated expression of the actuation stiffness can be written as
\begin{equation}
{K_{\varphi \varphi }} =  - \frac{{\partial {F_\varphi }}}{{\partial \varphi }} =  - \frac{1}{2}\sum\limits_{i = 1}^4 {\frac{{{\partial ^2}{C_{{\rm{i,TM}}}}}}{{\partial {\varphi ^2}}}} {V_{\rm{i}}}^2
\label{eq:4},
\end{equation}
where $\frac{{{\partial ^2}{C_{{\rm{i,TM}}}}}}{{\partial {\varphi ^2}}}$ means the second-order derivative of ${C_{{\rm{i,TM}}}}$ with respect to $\varphi$.

Combining assumptions (3) and (4) with equations~(\ref{eq:3}) and (\ref{eq:4}) yields the equation set for the needed actuation voltages in the case of $\varphi$-torque actuation:
\begin{equation}
\begin{cases}
 {F_x} = \frac{1}{2}\sum\limits_{i = 1}^4 {\frac{{\partial {C_{{\rm{i,TM}}}}}}{{\partial x}}} {V_{\rm{i}}}^2 = 0 \\
 {N_\varphi } = \frac{1}{2}\sum\limits_{i = 1}^4 {\frac{{\partial {C_{{\rm{i,TM}}}}}}{{\partial \varphi }}} {V_{\rm{i}}}^2 \\
 {K_{\varphi \varphi ,req}} =  - \frac{1}{2}\sum\limits_{i = 1}^4 {\frac{{{\partial ^2}{C_{{\rm{i,TM}}}}}}{{\partial {\varphi ^2}}}} {V_{\rm{i}}}^2 \\
 {V_{{\rm{TM}}}} = \sum\limits_{i = 1}^4 {{C_{{\rm{i,TM}}}}{V_i}}  = 0 \\
  \end{cases}.
\label{eq:5}
\end{equation}

\begin{figure}
\includegraphics[width=250pt]{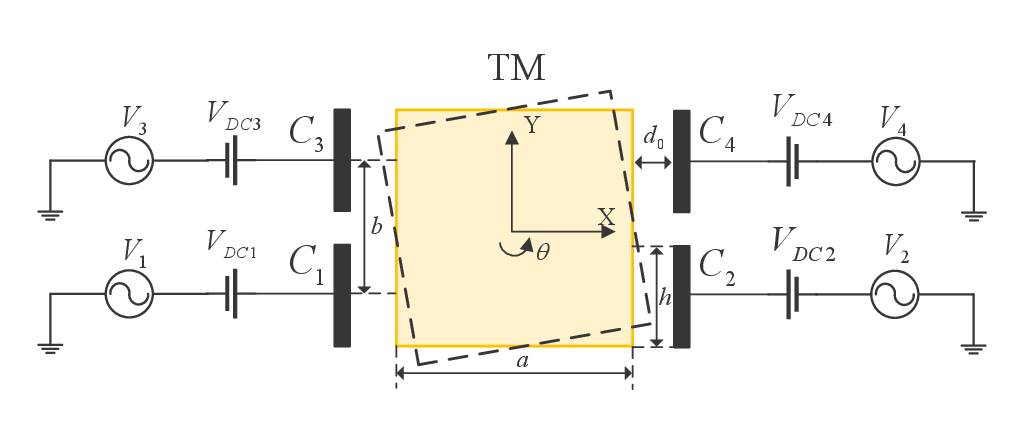}
\caption{\label{fig1} Schematic showing the geometry of a TM with four electrodes in the two-DOF system. The TM is indicated in yellow and the four x-direction electrodes are shown in black. $b$ is the distance between the centers of two electrodes on the same side, $h$ is the electrode width, $d_{0}$ is the nominal distance between the electrode and the TM, $V_{\rm{i}}$  represents the actuation voltage applied on one electrode, and  $V_{\rm{DCi}}$ is the DC bias voltage.}
\end{figure}

Figure~\ref{fig1} shows a schematic of the TM with four sensing and actuation electrodes. The solution of the four voltages can be expressed as
\begin{align}
 {V_1} &= \frac{1}{{\sqrt 2 }}\sqrt {\frac{1}{{\left| {\frac{{\partial {C_x}}}{{\partial \varphi }}} \right|}}} \sqrt {{N_\varphi } + \frac{{\left| {\frac{{\partial {C_x}}}{{\partial \varphi }}} \right|}}{{\left| {\frac{{{\partial ^2}{C_x}}}{{\partial {\varphi ^2}}}} \right|}}\left| {{K_{\varphi \varphi ,req}}} \right|} \label{eq:6} \\
 {V_2} &= \frac{1}{{\sqrt 2 }}\sqrt {\frac{1}{{\left| {\frac{{\partial {C_x}}}{{\partial \varphi }}} \right|}}} \sqrt { - {N_\varphi } + \frac{{\left| {\frac{{\partial {C_x}}}{{\partial \varphi }}} \right|}}{{\left| {\frac{{{\partial ^2}{C_x}}}{{\partial {\varphi ^2}}}} \right|}}\left| {{K_{\varphi \varphi ,req}}} \right|}  \label{eq:7}\\
 {V_3} &=  - \frac{1}{{\sqrt 2 }}\sqrt {\frac{1}{{\left| {\frac{{\partial {C_x}}}{{\partial \varphi }}} \right|}}} \sqrt { - {N_\varphi } + \frac{{\left| {\frac{{\partial {C_x}}}{{\partial \varphi }}} \right|}}{{\left| {\frac{{{\partial ^2}{C_x}}}{{\partial {\varphi ^2}}}} \right|}}\left| {{K_{\varphi \varphi ,req}}} \right|}  \label{eq:8}\\
 {V_4} &=  - \frac{1}{{\sqrt 2 }}\sqrt {\frac{1}{{\left| {\frac{{\partial {C_x}}}{{\partial \varphi }}} \right|}}} \sqrt {{N_\varphi } + \frac{{\left| {\frac{{\partial {C_x}}}{{\partial \varphi }}} \right|}}{{\left| {\frac{{{\partial ^2}{C_x}}}{{\partial {\varphi ^2}}}} \right|}}\left| {{K_{\varphi \varphi ,req}}} \right|}  \label{eq:9}
\end{align}
where $\left| {\frac{{\partial {C_x}}}{{\partial \varphi }}} \right|$ and $\left| {\frac{{{\partial ^2}{C_x}}}{{\partial {\varphi ^2}}}} \right|$
 represent the first- and second-order capacitance derivative:
\begin{align}
 \left| {\frac{{\partial {C_x}}}{{\partial \varphi }}} \right| &= \frac{{{C_0}b}}{{2{d_0}}} \label{eq:10}\\
 \left| {\frac{{{\partial ^2}{C_x}}}{{\partial {\varphi ^2}}}} \right| &= \frac{{3{b^2} + {h^2} + 3a{d_0}}}{{6{d_0}^2}}{C_0} \label{eq:11}
\end{align}
The solutions of equations~(\ref{eq:6})--(\ref{eq:9}) represent the amplitude of the actuation voltages; hence, they are renamed as
${V_1} = {u_{1\varphi }}$, ${V_2} = {u_{2\varphi }}$, ${V_3} = {u_{3\varphi }}$, and ${V_4} = {u_{4\varphi }}$. The maximum torque that can be exerted on the TM according to the required actuation stiffness is
\begin{equation}
{T_{\max }} = \frac{{\left| {\frac{{\partial {C_x}}}{{\partial \varphi }}} \right|}}{{\left| {\frac{{{\partial ^2}{C_x}}}{{\partial {\varphi ^2}}}} \right|}}\left| {{K_{\varphi \varphi ,req}}} \right|
\label{eq:12}
\end{equation}

The actuation voltages are applied to control the TM translation and rotation independently; hence, voltages on the electrodes must be orthogonal so that crosstalk between translational and rotational DOFs is avoided. Therefore, instead of steady DC voltages, either bipolar pulsed DC voltages or sinusoidal waveforms can be applied to the electrodes, both with zero average. In this study, the sinusoidal waveform scheme was chosen because it can be applied simultaneously in time, and orthogonality can be ensured using different frequencies. In our torsion pendulum system, we are concerned only with TM rotation. In this case, the voltages on each electrode are as follows:
\begin{equation}
\begin{aligned}
 {V_1} = {u_{1\varphi }}\sin 2\pi ft + {V_{DC1}}  \\
 {V_2} = {u_{2\varphi }}\cos 2\pi ft + {V_{DC2}} \\
 {V_3} = {u_{3\varphi }}\cos 2\pi ft + {V_{DC3}} \\
 {V_4} = {u_{4\varphi }}\sin 2\pi ft + {V_{DC4}}
\label{eq:13}
\end{aligned}
\end{equation}
where $f$ represents the frequency and ${V_{\rm{DCi}}}$ is the applied DC bias voltages to either compensate the stray potentials\cite{Antonucci2012, Apple2022} or bias the TM potential while charging or discharging\cite{Armano_2017, Armano2018}. According to the LPF actuation scheme\cite{Armano2020_actuation, Mance2012}, the frequencies should not be an integer factor of the 100-kHz sensing frequency, and their separation must be higher than 20 Hz to prevent the mixing of low frequencies in the actuation baseband. However, in this study, a 10-Hz frequency was chosen for the sinusoidal waveforms used in our experiment, for two reasons. First, because only one DOF was considered in the experiment, there is no crosstalk effect. Therefore, the frequency has no impact on control performance. Second, this approach avoids the need for massive storage space while conducting long-term experiments.

\subsection{\label{sec:level1} Analysis of TM Torque Noise}
To evaluate the influence factors on TM torque, the torque on the TM is expanded and written as
\begin{equation}
{N_\varphi } = \frac{1}{2}\left| {\frac{{\partial {C_x}}}{{\partial \varphi }}} \right|({V_2}^2 + {V_3}^2 - {V_1}^2 - {V_4}^2)
\label{eq:14}
\end{equation}
Substituting equation~(\ref{eq:13}) into equation~(\ref{eq:14}) and ignoring the AC terms (because their average torque contributions are zero) yields
\begin{equation}
\begin{split}
{N_\varphi} &= \frac{1}{4}\left| {\frac{{\partial {C_x}}}{{\partial \theta }}} \right|({u_{2\varphi }}^2 + {u_{3\varphi }}^2 - {u_{1\varphi }}^2 - {u_{4\varphi }}^2) \\
&+ \frac{1}{2}\left| {\frac{{\partial {C_x}}}{{\partial \theta }}} \right|({V_{DC2}}^2 + {V_{DC3}}^2 - {V_{DC1}}^2 - {V_{DC4}}^2)
\end{split}
\label{eq:15}
\end{equation}
From equation~(\ref{eq:15}), we can calculate the corresponding acceleration noise ${S_{{{N_\varphi}}}}$ due to the voltage amplitude noise
${S_{{u_{i\varphi }}}}$ and the DC actuation noise ${S_{{V_{DCi}}}}$. The total noise is given by
\begin{equation}
{S_{{N_\varphi }}}^{1/2} = \sqrt {\sum\limits_{i = 1}^4 {{S_{{u_{i\varphi }}}}{{(\frac{{\partial {N_{\rm{\varphi }}}}}{{\partial {u_{{\rm{i\varphi }}}}}})}^2}}  + \sum\limits_{i = 1}^4 {{S_{{V_{DCi}}}}{{(\frac{{\partial {N_{\rm{\varphi }}}}}{{{\partial _{{V_{DCi}}}}}})}^2}} }
\label{eq:16}
\end{equation}
where
\begin{eqnarray}
 \frac{{\partial {N_{\rm{\varphi }}}}}{{\partial {u_{1\varphi }}}} =  - \frac{{\partial {N_{\rm{\varphi }}}}}{{\partial {u_{4\varphi }}}} =  - \frac{1}{{2\sqrt 2 }}\sqrt {\left| {\frac{{\partial {C_x}}}{{\partial \varphi }}} \right|} \sqrt {{N_\varphi } + {T_{\max }}}  \label{eq:17}\\
 \frac{{\partial {N_{\rm{\varphi }}}}}{{\partial {u_{2\varphi }}}} =  - \frac{{\partial {N_{\rm{\varphi }}}}}{{\partial {u_{3\varphi }}}} = \frac{1}{{2\sqrt 2 }}\sqrt {\left| {\frac{{\partial {C_x}}}{{\partial \varphi }}} \right|} \sqrt { - {N_\varphi } + {T_{\max }}}  \label{eq:18}
\end{eqnarray}
Assuming no DC bias voltages are applied to the electrodes, i.e., $V_{\rm{DCi}}=0$, and the voltage amplitude noise ${S_{{u_{i\varphi }}}}$ is the same for all four electrodes, i.e., ${S_{{u_{1\varphi }}}} = {S_{{u_{2\varphi }}}} = {S_{{u_{3\varphi }}}} = {S_{{u_{4\varphi }}}} = {S_{{u_\varphi }}}$, then equation~(\ref{eq:16}) can be rewritten as
\begin{equation}
{S_{{N_\varphi }}}^{1/2} = {S_{{u_\varphi }}}^{1/2}\sqrt {\frac{1}{2}\left| {\frac{{\partial {C_x}}}{{\partial \varphi }}} \right|{T_{\max }}}
\label{eq:19}
\end{equation}
Further, by ignoring the AC feedback voltages applied to the electrodes and assuming that the DC actuation noise ${S_{{V_{DCi}}}}$ for all four electrodes is the same (that is, ${S_{{u_{1\varphi }}}} = {S_{{V_{DC1}}}} = {S_{{V_{DC1}}}} = {S_{{V_{DC1}}}} = {S_{{V_{DC}}}}$), the relationship between the torque noise and the DC actuation noise can be expressed as
\begin{equation}
{S_{{N_\varphi }}}^{1/2} = {S_{{V_{DC}}}}^{1/2}\left| {\frac{{\partial {C_x}}}{{\partial \varphi }}} \right|\sqrt {{V_{DC1}}^2 + {V_{DC2}}^2 + {V_{DC3}}^2 + {V_{DC4}}^2}
\label{eq:20}
\end{equation}
Therefore, the TM torque noise is related to the maximum torque $T_{\rm{max}}$ and the applied DC bias voltages ${V_{\rm{DCi}}}$. The effect of these two parameters on the TM torque noise by the torsion pendulum is discussed in Section \RNum{4}.

\section{Experiment Description}
\begin{figure}
\subfigure[Schematic drawing of torsion pendulum]{
\includegraphics[width=250pt]{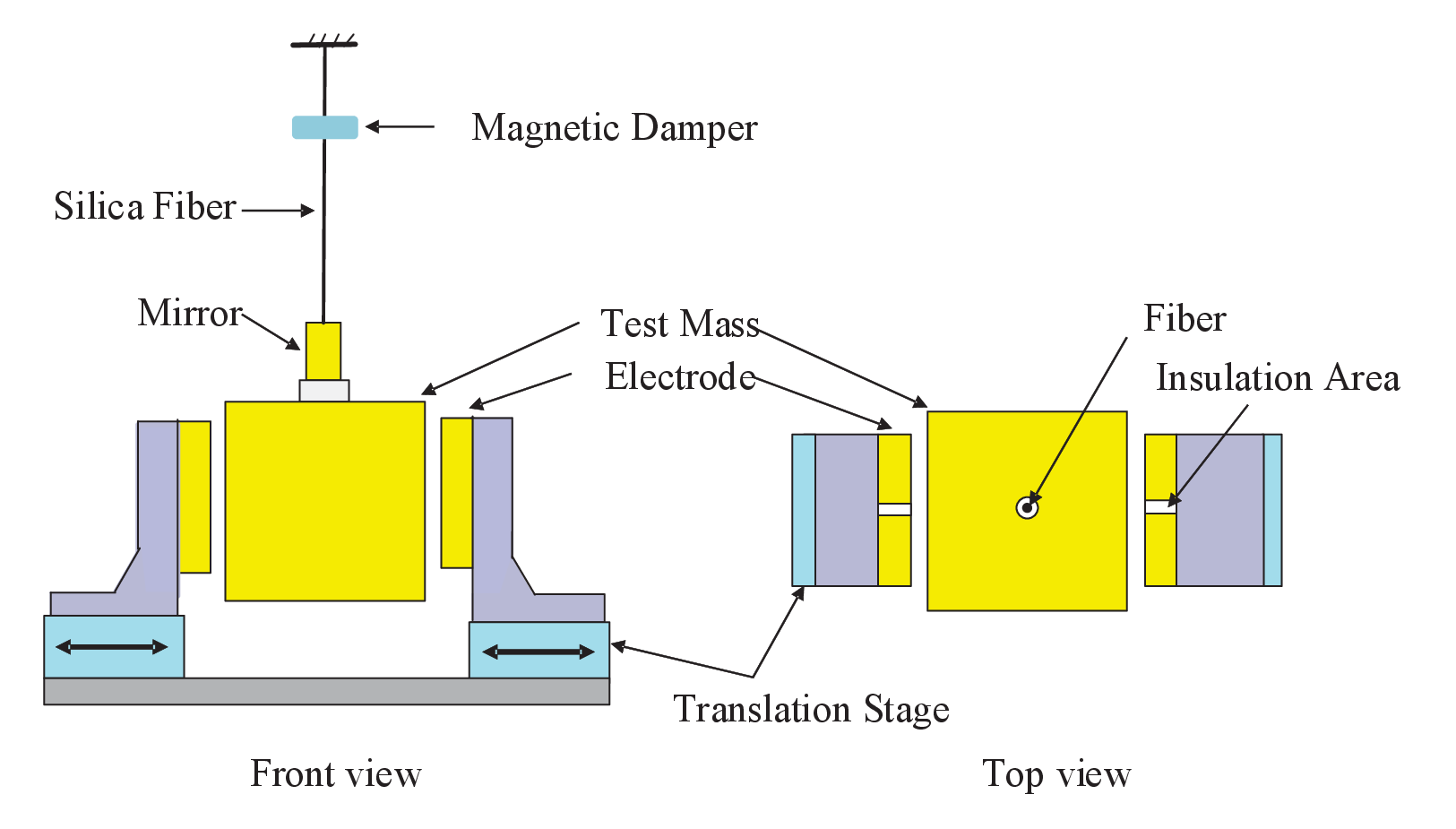}
}
\subfigure[Experimental setup]{
\includegraphics[width=250pt]{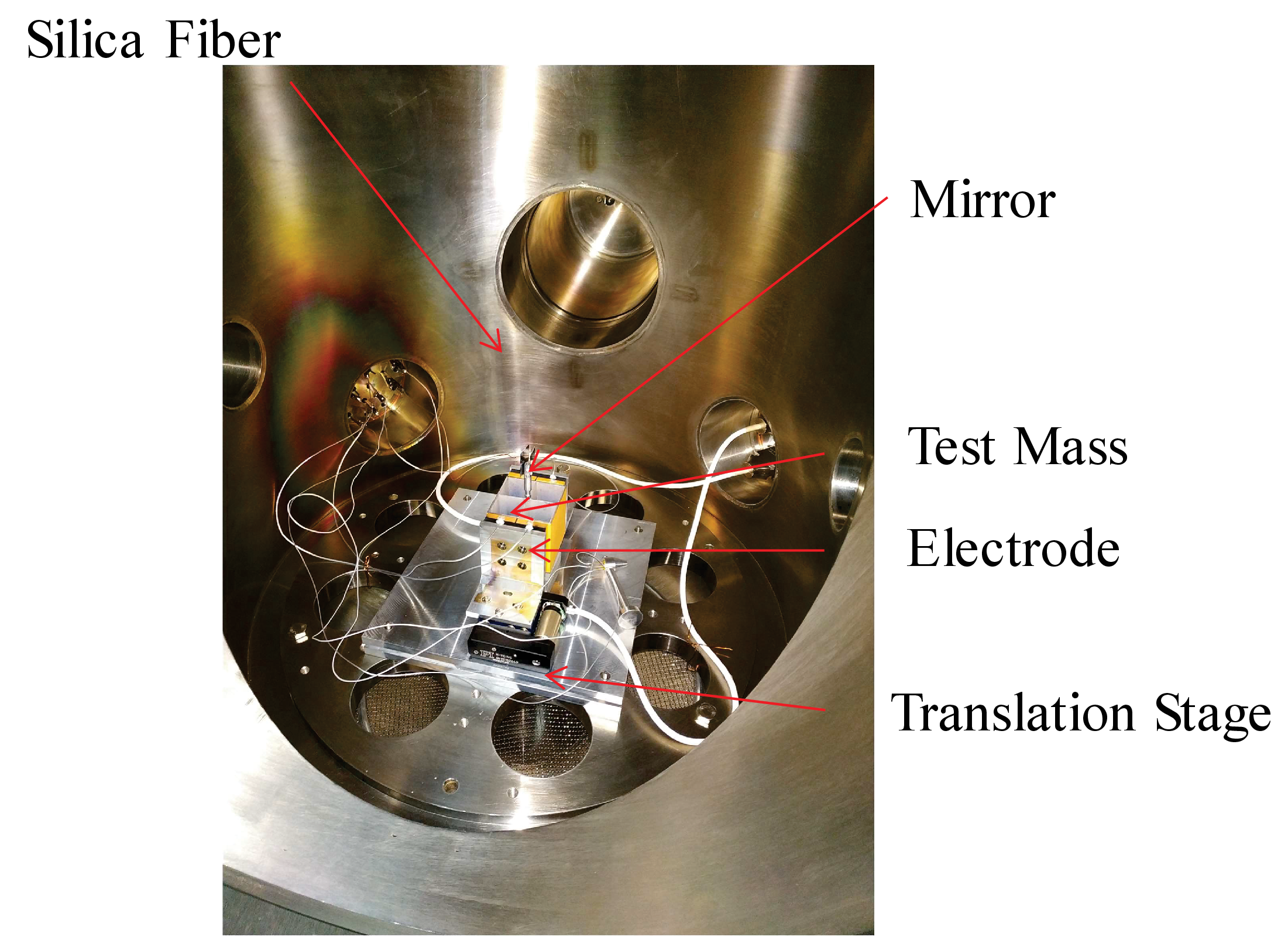}
}
\caption{\label{fig2} (a) Torsion pendulum. A pair of electrode plates were mounted symmetrically on two independent micro-displacement translation stages beside the TM. (b) Photographs of the experimental apparatus. The angle of the TM was monitored by an autocollimator.}
\end{figure}

A complete test on the ground for inertial sensors is highly difficult; hence, a single-mass torsion pendulum facility was developed in this study to obtain a “free-fall” status of the TM in the rotational direction to make low-frequency measurements with high torque sensitivity on the ground. Many precise measurements have been performed utilizing torsion pendulums. For example, a highly sensitive torsion pendulum system was used to measure and evaluate the performance of an inertial sensor\cite{Bassan2016, Carbone2003}, to measure the surface potential of a TM\cite{Pollack2008, Yin_2014}, to evaluate and demonstrate the feasibility of using ultraviolet LEDs to manage charge in gravitational-wave observatories\cite{Pollack_2010},  and to measure gas-damping variation\cite{Zhao2023}. A schematic drawing of our torsion pendulum is shown in Fig.~\ref{fig2}(a). The experimental setup is shown in Fig.~\ref{fig2}(b). A hollow TM was suspended with a silica fiber, and a pair of electrodes surrounding the TM were mounted on two independent translation stages to allow the gaps between the TM and electrodes to be adjusted.

The torsion pendulum system included a 50 mm $\times$ 50 mm $\times$ 50 mm hollow aluminum cube that was gold-plated to improve chemical stability, and a gold-coated mirror of dimensions 1.5 cm $\times$ 2.0 cm for autocollimator readout, made from glass ceramics. The cubic TM was suspended by a 895-mm-long, 40-$\mu$m-diameter germanium-and-bismuth-coated silica fiber from a magnetic damping stage, which was used to suppress its swing mode. The total suspended mass and moment of inertia were 58.6 g and 3.86 $\times$ 10$^{-5}$ kg m${^2}$, respectively.

\begin{figure}
\includegraphics[width=250pt]{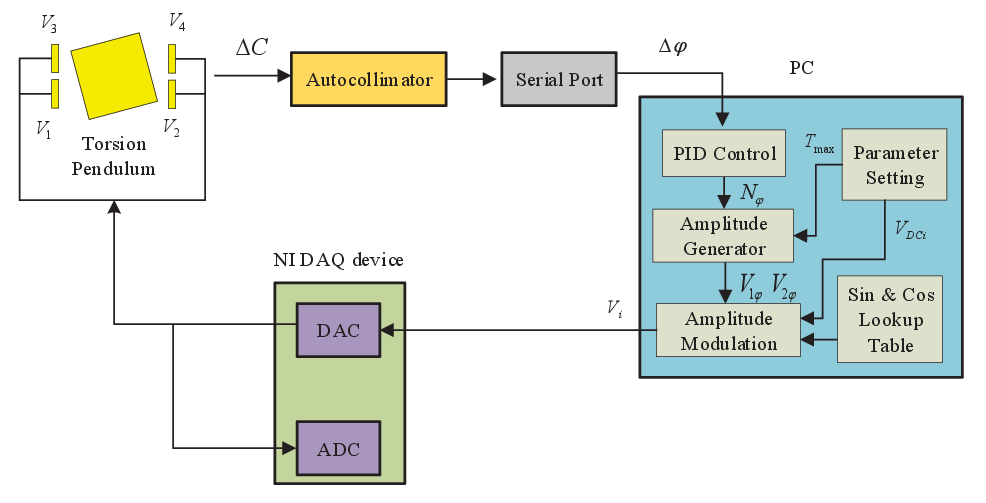}
\caption{\label{fig3} Closed-loop control system based on the torsion pendulum and electrostatic actuation.}
\end{figure}

Around the TM, a pair of gold-coated quartz electrodes of area 47 $\times$ 47 mm${^2}$ were fixed on two independent translation stages. Further, to control the rotation of the TM, the gold-coated layer of each electrode was divided vertically into two isolated parts. In this case, electrostatic actuation torque could be generated to drive the TM after applying corresponding voltages on each electrode. The torsion pendulum was installed in a vacuum chamber with a pressure 10${^{-5}}$ Pa. The free torsional oscillation period of our pendulum was approximately 490 s.

In this study, the charge on the TM was not considered because it was assumed to be eliminated by a charge management system in various space missions. Therefore, to reduce the influence of the electrostatic effect on the torsion pendulum, all devices that generate charged particles, such as the vacuum gauge and the residual-gas analyzer, were switched off during the measurement. Further, a charged-particle capture plate was placed between the torsion pendulum and the ion pump in the vacuum chamber to avoid random charging due to ionization.

To verify the actuation algorithm and evaluate the rotation control performance of the TM, a combined control system based on the torsion pendulum was designed (Fig.~\ref{fig3}). The rotational angle of the TM was detected by an autocollimator (ELCOMAT 3000) whose resolution was 0.1 $\mu$rad {Hz}$^{-1/2}$ at 1 mHz; the angle data was transmitted to a PC via serial ports. The PC included five modules: a proportional--integral--derivative (PID) controller, amplitude generator, amplitude modulation, lookup table, and parameter setting. The PID control module calculated the required electrostatic control torque for the TM from the input angle error in order to control the TM at a desired position. The amplitude generator module provided the amplitude of the sine and cosine waveforms utilizing the actuation algorithm mentioned in Section \RNum{2}. After receiving an amplitude command, the amplitude modulation module generated the expected AC feedback voltages according to the lookup table. The parameter setting module was used to set the values of various parameters, including maximum torque $T_{\rm{max}}$, DC bias voltage $V_{\rm{DCi}}$, and so on. Finally, AC feedback voltages were applied to the electrodes of the torsion pendulum system by the 16-bit digital-to-analog converter (DAC) in a NI DAQ device, with a maximum sampling rate of 3.3 MS/s.

\section{Results and Discussion}
\subsection{\label{sec:level2} Influence of parameters on TM torque noise}
\begin{figure}
 \subfigure[Relationship between amplitude of feedback voltages and the required electrostatic control torque]{
\includegraphics[width=250pt]{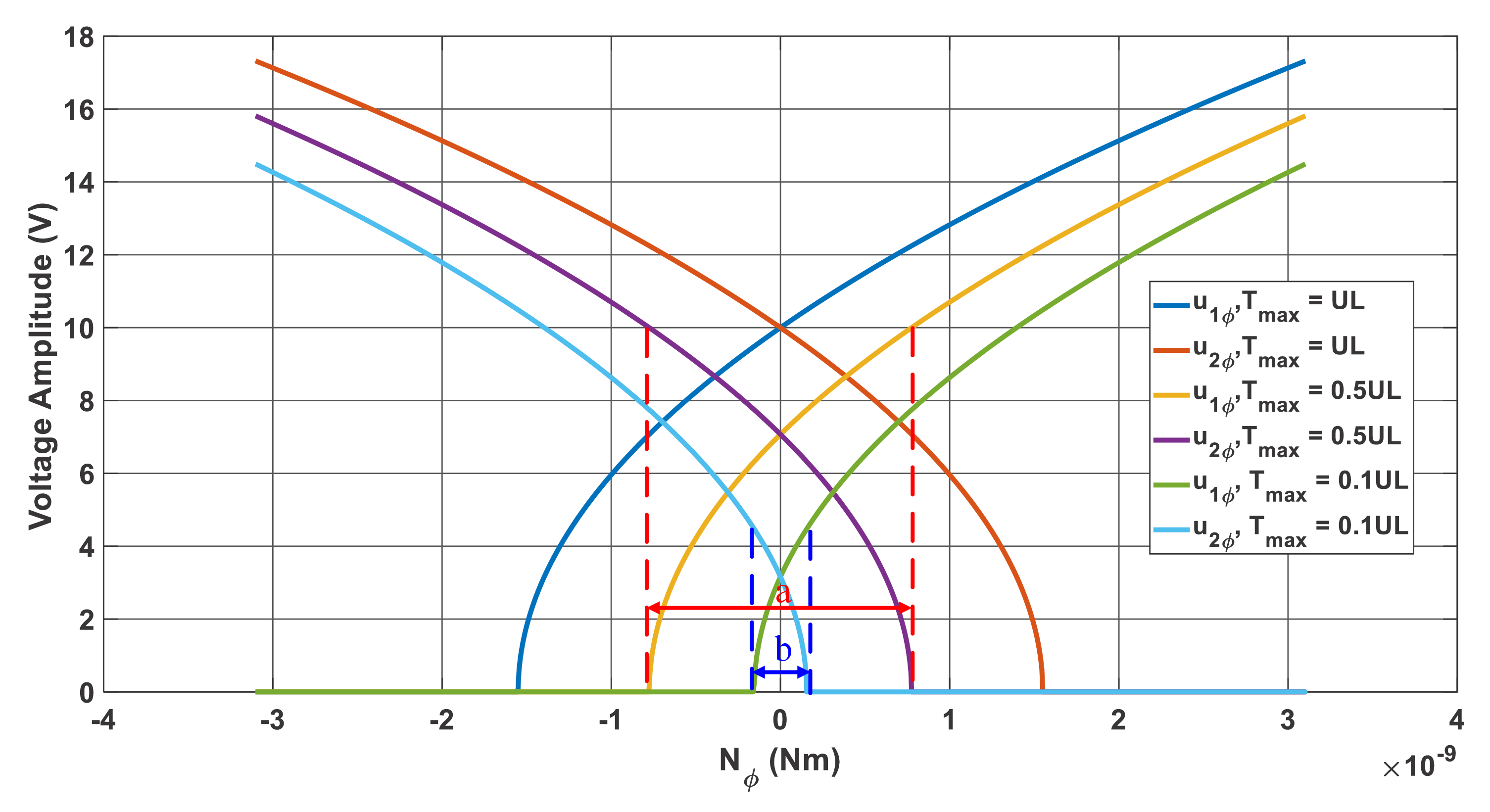}
}
\subfigure[Relationship between maximum voltage amplitude and maximum torque]{
\includegraphics[width=250pt]{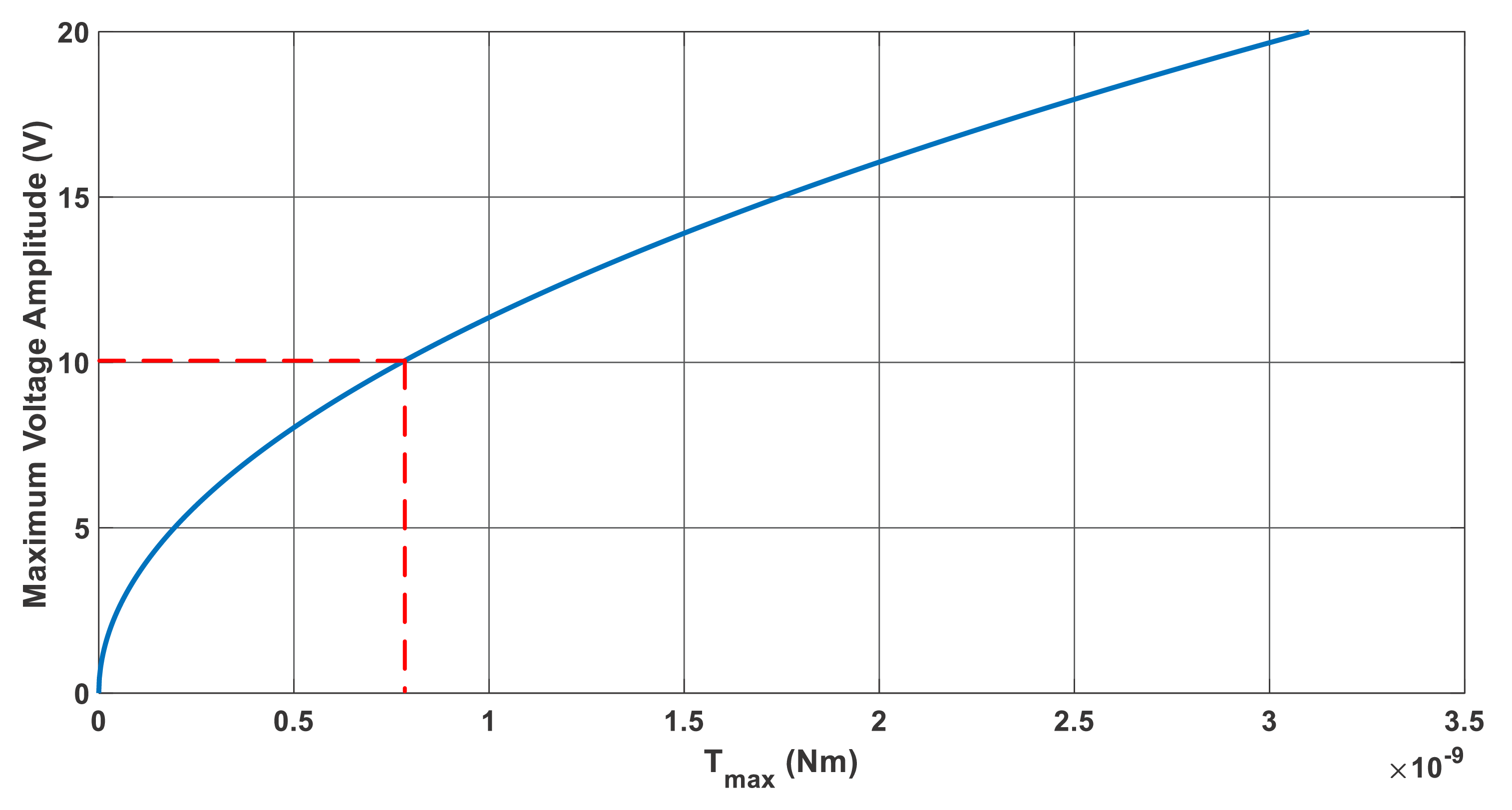}
}
\caption{\label{fig4} (a) Simulation results of feedback voltage amplitude varying with the required electrostatic control torque $N_{\varphi}$. (b) Simulation results of the maximum voltage amplitude ($\max \left| {{u_{1\varphi }} - {u_{2\varphi }}} \right|$) varying with $T_{\rm{max}}$. ${u_{1\varphi }}$ and ${u_{2\varphi }}$ represent the voltage amplitude for electrodes 1 and 2. Blue and red curves represent the variation of ${u_{1\varphi }}$ and ${u_{2\varphi }}$, respectively, in the case of $T_{max}$ = UL. The other curves have similar meaning. $a$ and $b$ indicate the adjustable ranges for $T_{\rm{max}}$ = 0.5UL and $T_{\rm{max}}$ = 0.1UL, respectively. Note that because ${u_{1\varphi }}={u_{4\varphi }}$ and ${u_{2\varphi }}={u_{3\varphi }}$, only ${u_{1\varphi }}$ and ${u_{2\varphi }}$ are shown.}
\end{figure}

To investigate the influence of maximum torque $T_{\rm{max}}$ on TM torque noise, it was necessary to determine the theoretical range of $T_{\rm{max}}$ in our experiment. Fig.~\ref{fig4} presents the simulation results. Fig.~\ref{fig4}(a) shows the relationship between feedback voltage amplitude and the required electrostatic control torque $N_{\varphi}$  with varying $T_{\rm{max}}$, based on equations~(\ref{eq:6})--(\ref{eq:9}). The maximum output voltage for the NI DAQ device used in this experiment was $\pm$10 V; hence, the upper limit value for $T_{\rm{max}}$ was approximately 1.55$\times$10$^{-9}$ Nm, which was called the UpperLimit (UL) value. Evidently, $N_{\varphi}$ can be zero only when $T_{\rm{max}}$ is selected as UL for keeping the feedback voltages below 10 V. In this case, the PID controller is ineffective because the adjustable range is zero. In contrast, for the other two situations ($T_{\rm{max}}$ = 0.5UL and $T_{\rm{max}}$ = 0.1UL), the PID controller can work properly because the adjustable ranges of $N_{\varphi}$ are $\pm$7.75$\times$10$^{-10}$ Nm and $\pm$1.55$\times$10$^{-10}$ Nm, which are indicated as $a$ and $b$, respectively, in Fig.~\ref{fig4}(a). Fig.~\ref{fig4}(b) presents the relationship between maximum voltage amplitude ($\max \left| {{u_{1\varphi }} - {u_{2\varphi }}} \right|$) and maximum torque $T_{\rm{max}}$. As indicated in the figure, the upper limit for the value of maximum torque was chosen as approximately $\pm$7.75$\times$10$^{-10}$ Nm in this experiment to keep the maximum voltage amplitude below 10 V. This value of maximum torque is called MAXVALUE in the following discussion of experimental results.

\begin{figure}
 \centering
\includegraphics[width=250pt]{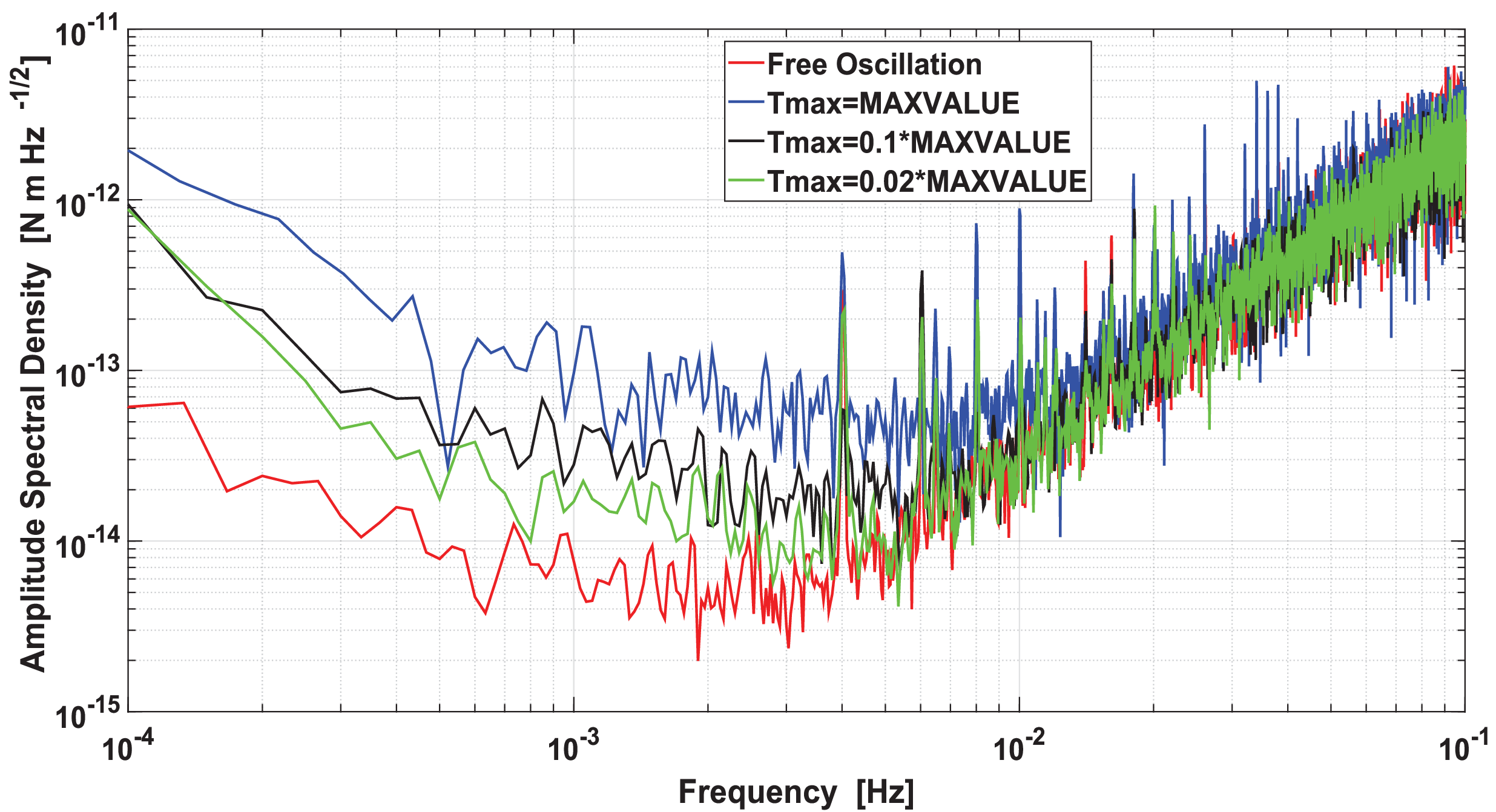}
\caption{\label{fig5} Amplitude spectral density (ASD) of the torque on the TM in our torsion pendulum with different maximum torques. The red, blue, black and green curves represent the maximum torque values of zero (free oscillation), MAXVALUE, 0.1$\times$MAXVALUE, and 0.02$\times$MAXVALUE, respectively. The measurement shown represents 60000 s of data; the ASD was calculated by windowing the signal with a 23000-s-long Hanning window with 50${\rm{\% }}$ overlap.}
\end{figure}

To investigate the influence of the maximum torque $T_{\rm{max}}$ on TM torque noise, a series of sine and cosine waveforms with different amplitudes generated by the amplitude modulation module were applied to the electrodes. Then, the electrostatic forces on the TM produced by electric fields were derived from the torque acting on the pendulum. This in turn was derived from the angular readout and the known geometry and dynamics of the pendulum. Fig.~\ref{fig5} shows an example of the amplitude spectral density of the pendulum’s sensitivity to the TM torque noise, representating the performance achieved during the measurements with different values of $T_{\rm{max}}$ as reported here. The red curve indicates the measured torque sensitivity of the free torsion pendulum, which was approximately $\pm$8$\times$10$^{-15}$ Nm/Hz$^{1/2}$ at 1 mHz, limited by direct torques on the TM and angular readout performance. Other curves indicate the corresponding torque sensitivity with different $T_{\rm{max}}$ values. The maximum torque noise was observed to reach $10^{-{13}}$ Nm/{Hz}$^{1/2}$ at 1 mHz (for $T_{\rm{max}}$ = MAXVALUE), which is 10 times larger than the torque sensitivity of the torque pendulum; reducing the value of $T_{\rm{max}}$ can effectively reduce the torque noise. In addition, the ratios of different noise signals at 1 mHz match the variation of $\sqrt {T_{\rm{max}}}$, which agrees with Equation~(\ref{eq:19}). Note that the PID controller was disabled and the system was an open-loop system in this case.

\begin{figure}
 \centering
\includegraphics[width=250pt]{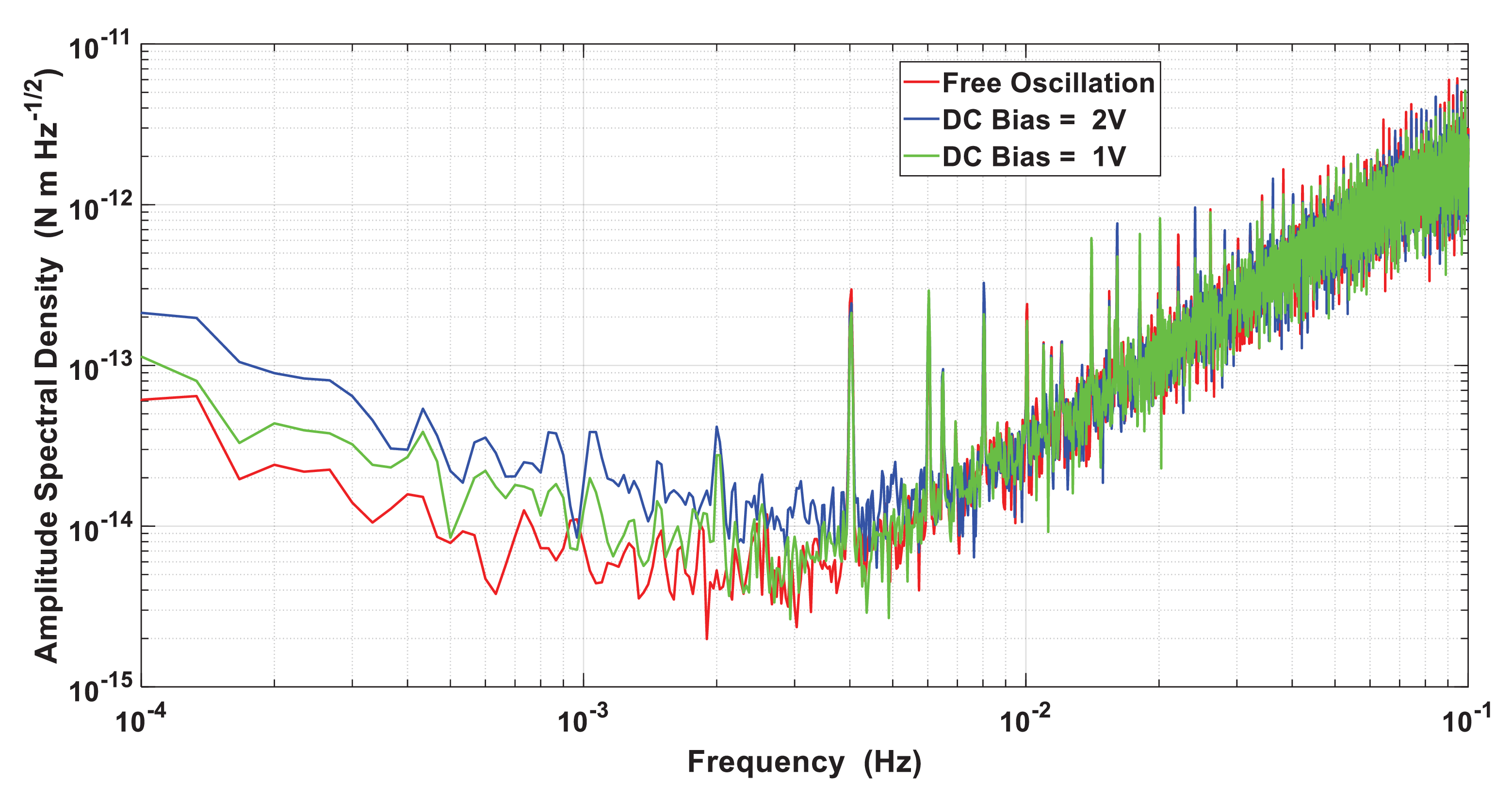}
\caption{\label{fig6} ASD of the torque on the TM in our torsion pendulum with different DC bias voltages. The red, blue, and green curves represent the DC bias voltages of 0 V (free oscillation), 2 V, and 1 V, respectively.}
\end{figure}

Further, to test the influence of the DC bias voltages on TM torque noise, only DC voltages were applied to the four electrodes. A number of positive and negative DC voltages were applied on electrodes 1 to 4 to avoid the induced potential on the TM. DC bias voltages were assumed to have an impact on the torque noise, particularly when the DC voltage was greater than 1 V (Fig.~\ref{fig6}). As mentioned previously, the DC bias was to compensate for the stray potential on electrodes or to bias the TM potential while charging or discharging. For the former, because typical stray potential is approximately tens to hundreds of millivolts, DC bias at this level cannot affect torque noise. For the latter, because the TM charging or discharging process is rapidly operated in fast charge mode by the charge management system, the acceleration data are ignored during this time. Therefore, the influence of DC actuation noise on the TM torque noise was neglected in this experiment.

\subsection{\label{sec:level3}  TM Control Performance}
 \begin{figure}
 \subfigure[Angular readout of TM]{
\includegraphics[width=250pt]{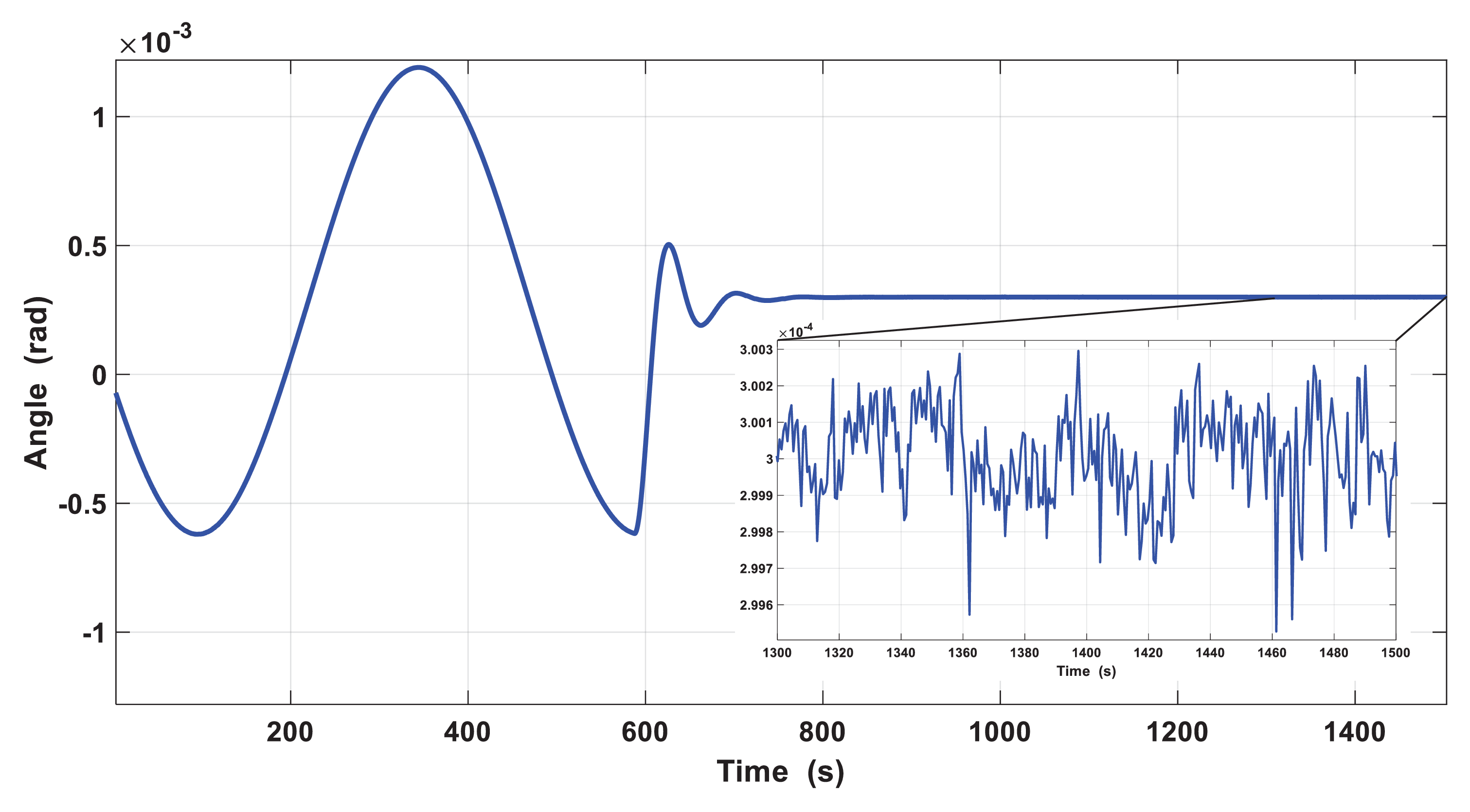}
}
\subfigure[Amplitude of feedback voltages]{
\includegraphics[width=250pt]{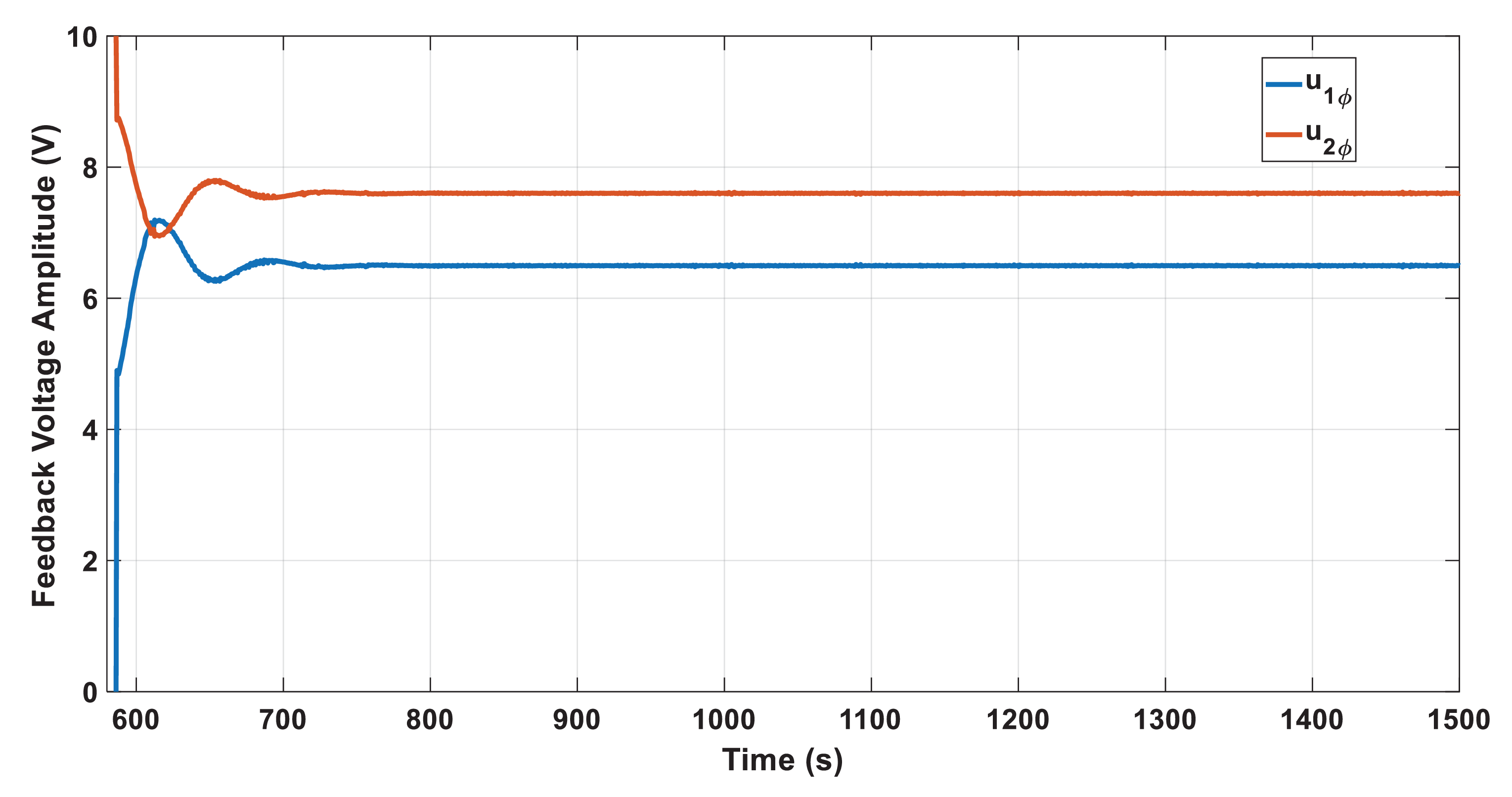}
}
\subfigure[Feedback voltages applied on four electrodes]{
\includegraphics[width=250pt]{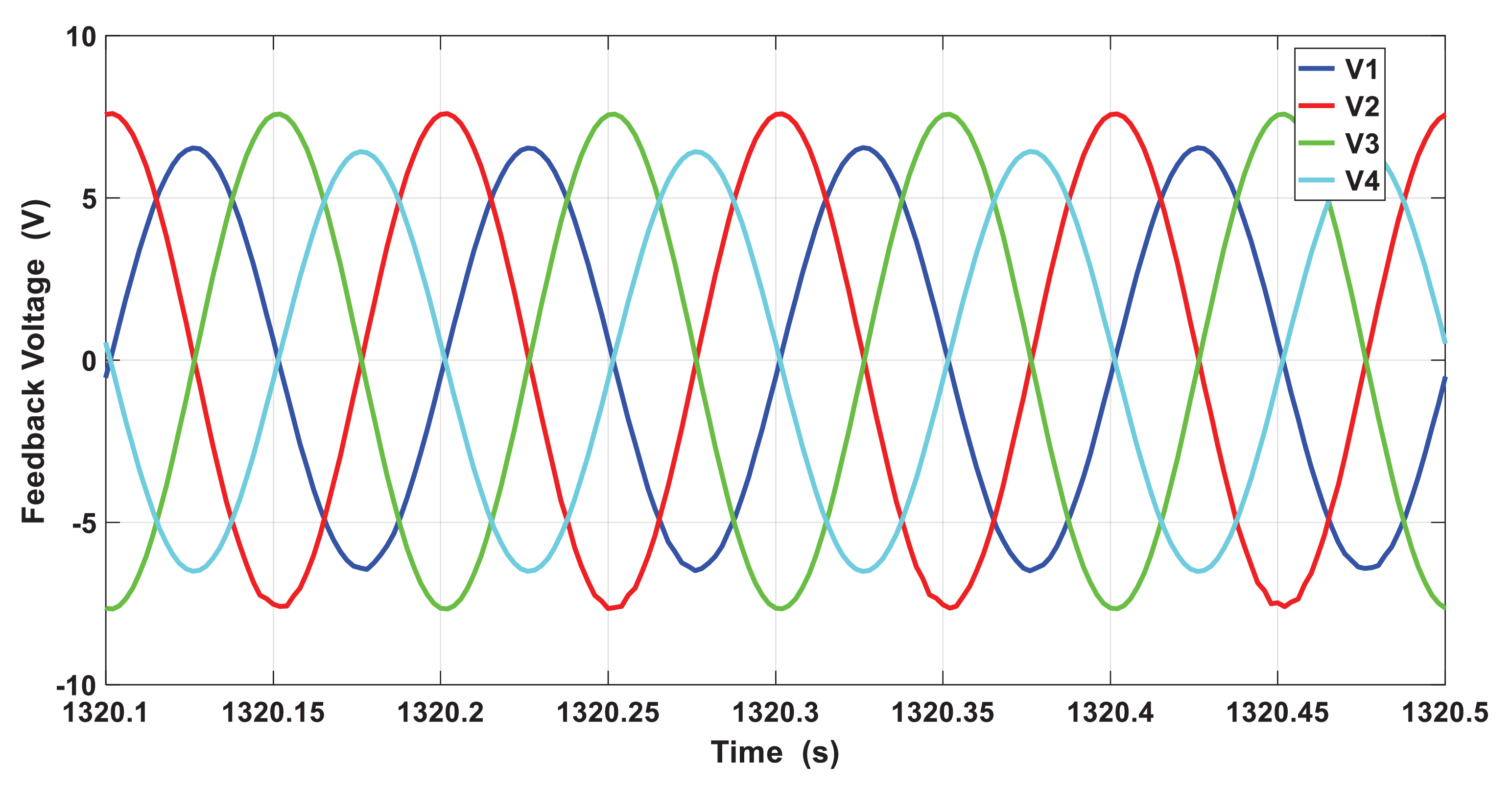}
}
\subfigure[Amplitude spectral density (ASD) of the feedback voltages]{
\includegraphics[width=250pt]{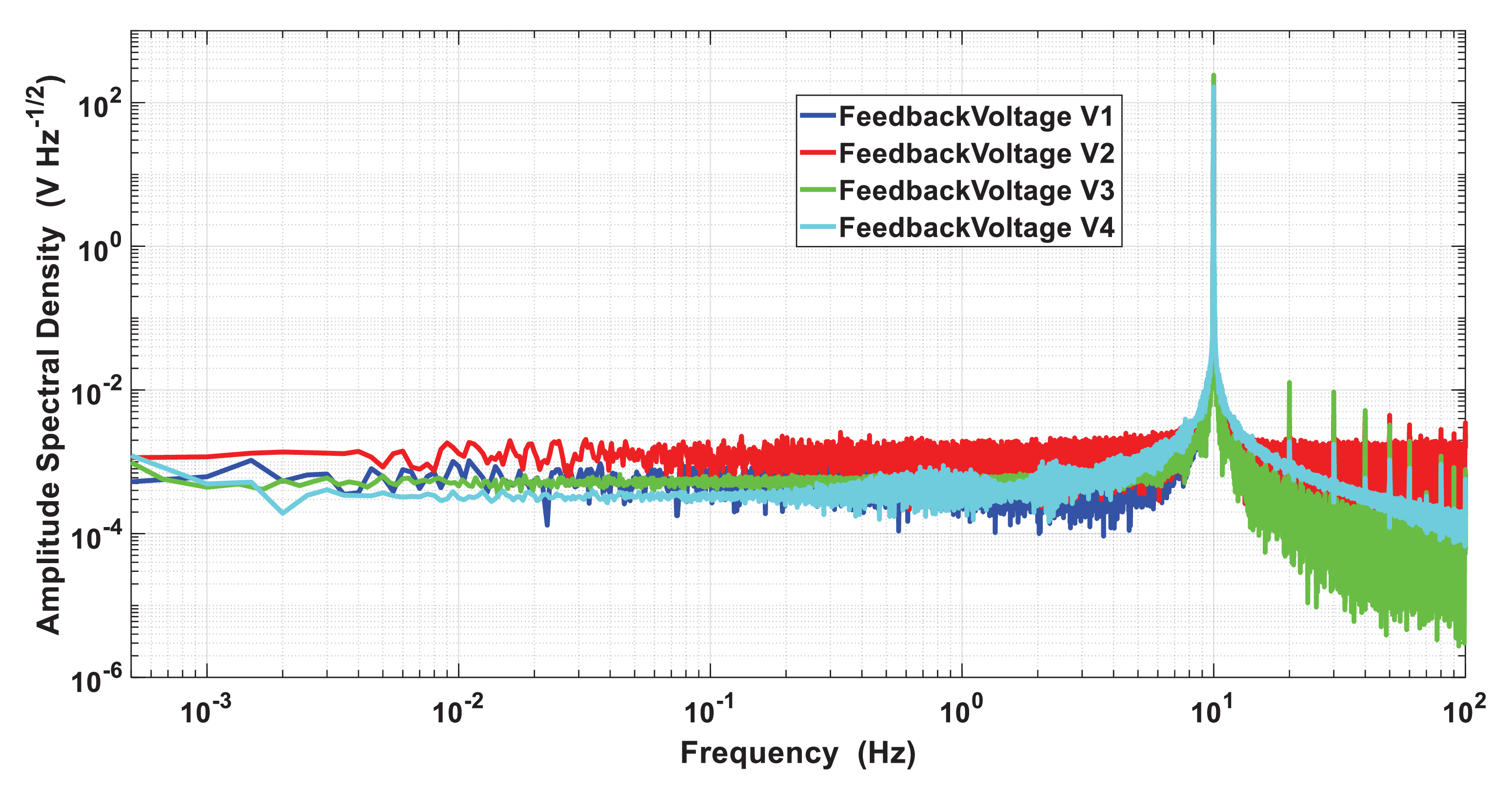}
}
\caption{\label{fig7} Dynamic performance and relevant parameters of the closed-loop system while controlling the TM. (a) Angular readout of the TM, showing the dynamic performance. (b) Amplitude of the feedback voltages. (c) Feedback voltages applied on the four electrodes. (d) ASD of the feedback voltages on the electrodes.}
\end{figure}

The overall stable closed-loop system was constructed and its performance was evaluated. The experimental process was as follows. First, the PID controller was disabled and the four electrodes were shorted to ground to let the TM oscillate freely. Then, the PID controller was enabled, and the required AC feedback voltages were calculated automatically and applied to the electrodes. Several groups of control parameters were available for the PID controller, and could be switched from one to another during the experiment to optimize control performance. Finally, the relevant parameters were recorded and the performance was evaluated.

Fig.~\ref{fig7} shows the dynamic performance and relevant parameters of the closed-loop system while controlling. The TM without control maintained a state of free oscillation with an amplitude of approximately 1.8 mrad (Fig.~\ref{fig7}(a)). Then, at approximately 590 s, the controller was enabled and the amplitude of oscillation decreased gradually to the desired equilibrium position, 0.3 mrad. The stable tracking error was at a level of approximately 10$^{-7}$. Fig.~\ref{fig7}(b), (c), and (d) represent the amplitude, applied feedback voltages, and ASD of the voltages. The amplitudes calculated by the actuation algorithm were approximately 7.6 V and 6.5 V, respectively, in the case of $T_{\rm{max}}$ = MAXVALUE. The measured output voltage sensitivity of the DAC integrated in the NI DAQ device was approximately 8$\times$10$^{-4}$ V/Hz$^{1/2}$ for channels 1, 3, and 4, and 10$^{-3}$ V/Hz$^{1/2}$ for channel 2, at 1 mHz. Notably, the measurements were repeated several times and this sensitivity mismatching was consistently observed.

The small adjustable range may not provide sufficient driving force while controlling the TM rotation, resulting in unstable control performance, such as slow dynamic response, large overshoots, and low steady-state accuracy. Therefore, the TM rotation control performance was investigated for different maximum torque values. Several measurements of the combination closed-loop system were conducted by varying the maximum torque values while tracking a series of step-input commands of different magnitudes. The step-input commands were divided into two parts: large steps and small steps. Large steps were used to simulate the working condition when the TM was released by the caging mechanism, and they were set as a vector signal of the sequence stair, $\vec \alpha$=[0, 3, 6, 3, 0] mrad. Small steps were adopted to simulate the fine-tuning process after the TM was captured by the control system; they were also set as a vector signal of the sequence stair, $\vec \beta$ =[0.3, 0.6, 0.9, 0.6, 0.3] mrad. Each step lasted 500 s so that the total control time was 2500 s.

 \begin{figure}
 \centering
\includegraphics[width=250pt]{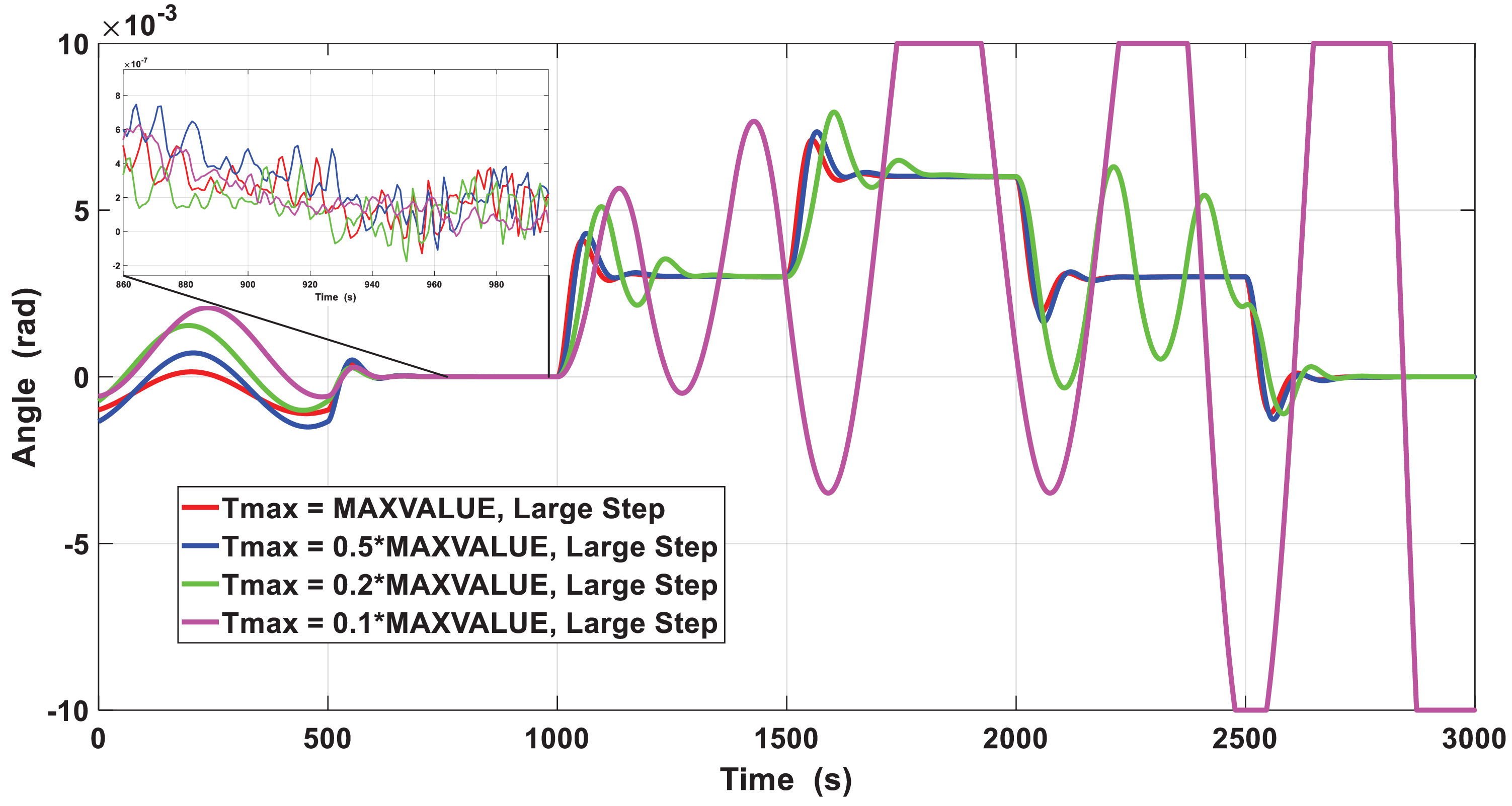}
\caption{\label{fig8} Responses of the overall closed-loop system and the tracking errors of the first step, demonstrated with large steps. Red, blue, green, and purple curves represent the maximum torque values of MAXVALUE, 0.5$\times$MAXVALUE, 0.2$\times$MAXVALUE, and 0.1$\times$MAXVALUE, respectively.}
\end{figure}

 \begin{figure}
 \centering
\includegraphics[width=250pt]{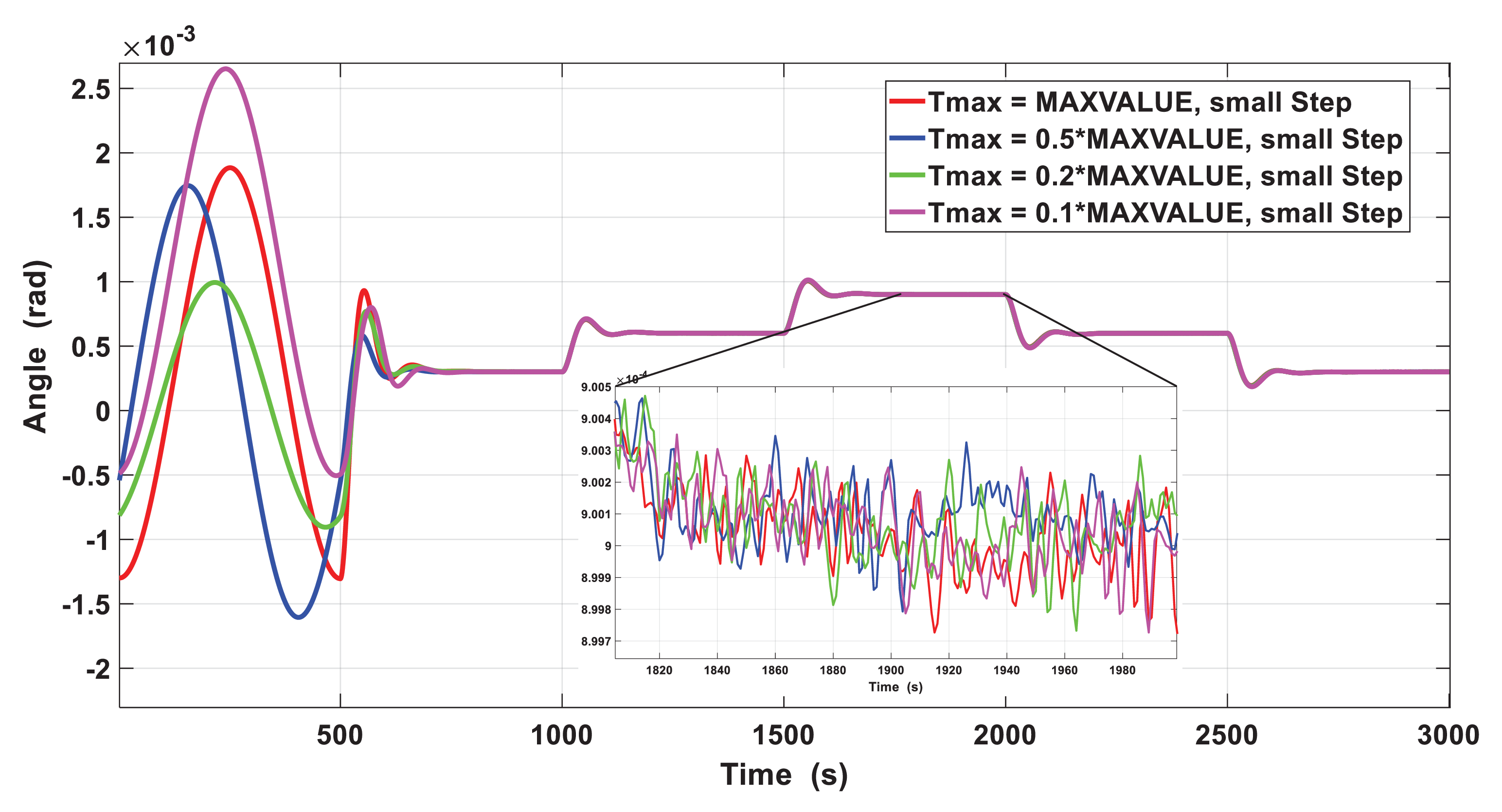}
\caption{\label{fig9} Responses of the overall closed-loop system and the tracking errors of the first step, demonstrated with small steps. Red, blue, green, and purple curves represent the maximum torque values of MAXVALUE, 0.5$\times$MAXVALUE, 0.2$\times$MAXVALUE, and 0.1$\times$MAXVALUE, respectively.}
\end{figure}

The responses of the overall closed-loop system and the tracking errors of the first step are shown in Fig.~\ref{fig8} with large steps $\vec \alpha$. The results show that the system was unable to track the input command when the maximum torque was chosen as 0.2$\times$MAXVALUE and  0.1$\times$MAXVALUE; the tracking errors increased gradually over time. Conversely, the TM in the other two cases followed the expected position well. These results show that the maximum torque should be greater than 1.55$\times$10$^{-10}$ Nm to provide sufficient driving force to improve tracking performance and robustness for large-step conditions. The tracking errors in the first cycle are also shown (magnified view). The experiments were also conducted with small steps $\vec \beta$, as shown in Fig.~\ref{fig9}. In those conditions, all measurements provided adequate closed-loop tracking performance.

Overall, the combined closed-loop control system could verify the actuation algorithm, and the control performance was evaluated with different maximum torques under different working conditions of inertial sensors (large steps and small steps). These results can provide a reference for optimizing TM control performance under different conditions for space missions.

\section{Conclusion}
In this study, we developed a torsion pendulum to investigate the influence of various parameters in the electrostatic actuation system on TM torque noise. We designed a combined system of the torsion pendulum and actuation modules to assess TM control performance for high-precision space-based missions. The results show that the additional torque noise introduced by the parameters in an actuation system can reach $10^{-{13}}$ Nm/{Hz}$^{1/2}$ at 1 mHz. The stable tracking error for the closed-loop system was approximately 10$^{-7}$, indicating that the combined system achieves good tracking performance and robustness for TM rotation control under various inertial sensor conditions.

First, the mathematical model of TM actuation algorithm and TM torque noise were derived, showing that the TM torque noise is related to the maximum torque $T_{\rm{max}}$ and the applied DC bias voltages $V_{\rm{DCi}}$. Then, a series of sine and cosine waveforms with different amplitudes were applied to the electrodes to observe the influence of these two parameters on TM torque noise. Finally, a combined closed-loop control system based on the torsion pendulum and actuation modules was constructed, and the responses of the overall closed-loop system as well as the tracking errors were evaluated with different $T_{\rm{max}}$ values. The results indicate that the maximum torque and DC bias voltages introduce additional torque noise. The maximum noise can reach up to $10^{-{13}}$ Nm/{Hz}$^{1/2}$ at 1 mHz. The stable tracking error for the closed-loop system was found to be approximately 10$^{-7}$, indicating that the combined system achieves good tracking performance and robustness for TM rotation control in different working conditions. Notably, the voltage amplitude noise ${S_{{u_{i\varphi }}}}$ and the DC actuation noise ${S_{{V_{\rm{DCi}}}}}$ observed in this study were relatively higher than that in regular space missions (approximately 100$\sim$200 $\mu$V/Hz$^{1/2}$) because it was limited by the NI instrument. Therefore, the selection range of these two parameters should be wider in the case of in-orbit applications.

The aim of this work was to investigate the acceleration noise caused by the actuation system and evaluate the tracking performance for the TM position control, which is a key technology in space-borne experiments. The analysis of the mathematical model and experimental method described here provides effective test methods and performance evaluation methods for actuation systems, which is applicable to the development of actuation systems for numerous space gravitational missions.

\begin{acknowledgments}
This work was supported by the National Natural Science Foundation of China under Grant number 12105250.
\end{acknowledgments}

\bibliography{arxiv_version}

\end{document}